\documentclass[fleqn,usenatbib]{mnras}

\usepackage[T1]{fontenc}

\DeclareRobustCommand{\VAN}[3]{#2}
\let\VANthebibliography\thebibliography
\def\thebibliography{\DeclareRobustCommand{\VAN}[3]{##3}\VANthebibliography}

\usepackage{graphicx}	

\title[Decameter radio emission from Jupiter related to Ganymede]{On quasi-periodic decameter radio emission from Jupiter related to Ganymede}

\author[V. E. Shaposhnikov and V. V. Zaitsev]{
V.E. Shaposhnikov,$^{1,2}$\thanks{E-mail: sh130@ipfran.ru}
V.V. Zaitsev,$^{1}$
\\
$^{1}$Institute of Applied Physics of the Russian Academy of Sciences, Nizhny Novgorod,
Russia\\
$^{3}$High School of Economics, National Research University, Nizhny Novgorod Branch, Nizhny Novgorod, Russia}

\pubyear{2025}

\begin{document}
\label{firstpage}
\pagerange{\pageref{firstpage}--\pageref{lastpage}}
\maketitle

\begin{abstract}

The paper discusses the possibility of implementing a plasma generation mechanism in a source of decameter radio emission associated with Ganymede and an explanation based on this mechanism of forming quasi-periodic sequences of bursts of this radiation. According to the discussed model, the registered quasi-periodic sequences of radiation pulses are a consequence of implementing in the source a pulsating mode of plasma wave conversion into extraordinary electromagnetic waves with a small refractive index. The negative frequency drift of the radiation observed in the frequency-time spectrogram is due to the group delay of waves with a small refractive index and the dispersion of the medium. Based on the plasma model, estimates of the plasma parameters in the generation region are obtained, which are in agreement with the data obtained as a result of satellite measurements.

\end{abstract}



\section{Introduction}

Decameter radio emission from Jupiter is one of the most interesting objects of study in the Solar System. Long-term observations of this radiation have revealed a rich frequency-time structure of its spectrum. Particular attention has usually been paid to short-lived bursts - S-bursts, the appearance of which is caused by the interaction of the satellite Io with the planet's magnetosphere \citep[see, for example,][and the literature cited there]{Riihimaa(rcfggll)(1981),Zarka(2007),Panchenko(prrbzlskmfs)(2018)} . In the dynamic spectrum, these bursts often form quasi-periodic sequences of negatively drifting radiation pulses. As it turned out, such radiation is not the prerogative of the satellite Io alone. \citet{Mauduit(mzlh)(2023)} report the discovery of quasi-periodic sequences of decameter S-bursts associated with the satellite Ganymede and the auroral regions of Jupiter. Figure 1 shows examples of dynamic spectra with quasi-periodic frequency-drifting bursts of decameter radio emission associated with Ganymede.
 \begin{figure}
	\includegraphics[width=\columnwidth]{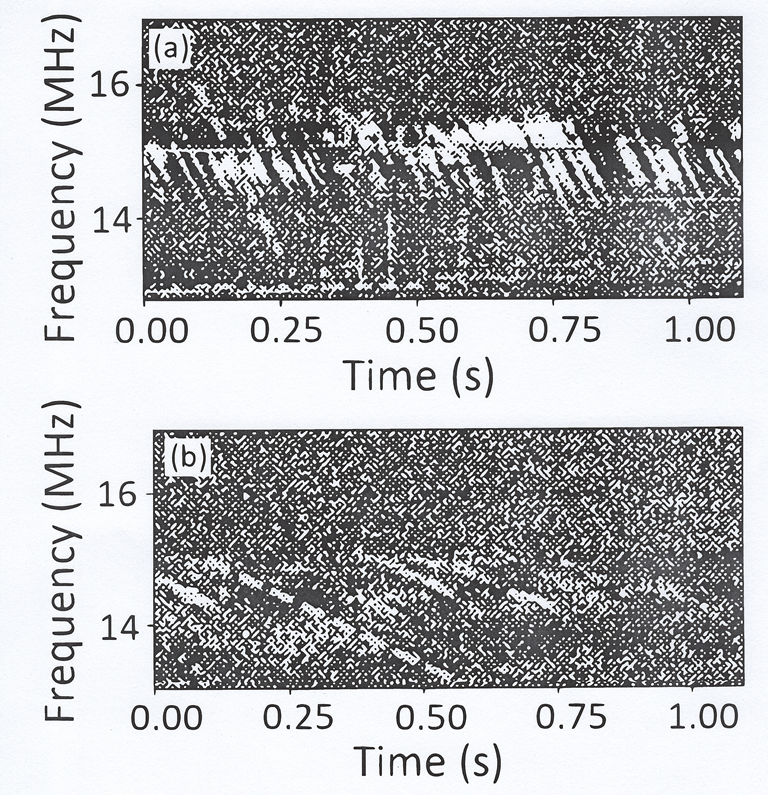}
    \caption{Examples of dynamic spectra with quasi-periodic sequences of frequency-drifting bursts of decameter radio emission from Jupiter related to Ganymede. Sequences with a period of 31 ms and a frequency drift rate of -16.9 MHz/s (a) and a  period of 38 ms and a frequency drift rate of -3.36 MHz/s (b). Figures adapted from Figure 2 in \citet{Mauduit(mzlh)(2023)}.}
    \label{fig1}
\end{figure}

The most widely discussed mechanism for generating decameter radio emission from Jupiter is the electron cyclotron maser (ECM), caused by weakly relativistic, on the order of several tens of keV, electrons with a nonequilibrium velocity distribution function \citep{Wu(wl)(1979),Louarn(1992),Zaitsev(zs)(1994)}. The use of this mechanism made it possible to explain the main properties of L-bursts of decameter radio emission: relation to the local electron cyclotron frequency, the high brightness temperature, and the thin walled hollow cone beamed radio emission.
The electron cyclotron maser has also been proposed as a mechanism for generating radio emission associated with  satellites Io and Ganymede
 \citep[see, for example,][]{Hess(hzm)(2007),Hess(hmz)(2007),Mauduit(mzlh)(2023)}. Like the L-emission, these bursts have a high brightness temperature and  the thin walled hollow cone beamed radio emission, and their frequency is close to the local electron cyclotron frequency. However, unlike the L-emission, the S-bursts have a rich time-frequency structure. In particular, these bursts appear as emission bands rapidly drifting toward lower frequencies in the dynamic spectrum. In some cases, these bursts are observed as quasi-periodic pulse trains \citep{Flagg(fkl)(1976),Mauduit(mzlh)(2023)}.

In the framework of the S-radiation source model based on the electron cyclotron maser, the negative frequency drift of the radiation is usually associated with the drift of the emitting electrons or the entire generation region along the magnetic field lines \citep{Ellis(1965),Zarka(1998),Mauduit(mzlh)(2023)}. The observed values of the frequency drift velocity require electrons with an energy of several tens of keV. For example, the frequency drift values of the S-bursts induced by Ganymede are concentrated mainly in two intervals: $-(14-22) \mbox{ MHz/s}$  and  $-(3-7) \mbox{ MHz/s}$, which corresponds to an energy of emitting electrons of $2-10 \mbox{ keV}$ and $0.1-1 \mbox{ keV}$ , respectively \citep{Mauduit(mzlh)(2023)}. It should be noted here that ECM is an inefficient radiation source, with nonequilibrium electron energies $<0.5 \mbox{ keV}$ \citep{Wong(wkw)(1989)}. This creates certain difficulties for the ECM interpretation of bursts with a small frequency drift.

The existence of quasi-periodic sequences of S-bursts is associated with the pulsating mode of the acceleration mechanism. It is assumed that an Alfven resonator is realized in the ionosphere of Jupiter. The Alfven wave coming from the satellite excites oscillations in the resonator in the frequency range containing the repetition frequencies of S-bursts \citep{Su(sjebpdl)(2006),Ergun(esabdls)(2006)}. This coincidence served as the basis for the assumption that the electric field of Alfven oscillations in the ionospheric Alfven resonator accelerates electrons to the required energies and forms pulses of energetic electrons. The pulses of energetic electron are the reason for the appearance of quasi-periodic sequences of S-bursts \citep{Ergun(esabdls)(2006),Hess(hmz)(2007)}.

As observations show, the indicated sequences contain a large number of individual S-bursts, of practically the same intensity \citep{Mauduit(mzlh)(2023)}. If we accept that the acceleration of the emitting electrons is caused by the electric field of the Alfven oscillations in the ionospheric resonator, this means that this resonator is high-quality and contains a large number of oscillations of approximately equal amplitude. At the same time, the presence of absorption, energy loss during electron acceleration and the imperfection of the resonator walls reduce the quality factor, which leads to a fairly rapid attenuation of the Alfven oscillations \cite[see, in this regard,][]{Lysak(ll)(1996)}.
It should be noted here that numerical MHD modeling (applied to terrestrial conditions) of the excitation of Alfven oscillations in the resonator by Alfven pulse incident on the ionosphere and the acceleration of electrons by the electric field of these oscillations showed that the number of pulses of accelerated electrons is small, and the energy of electrons in a pulse decreases with time \citep{Chaston(cbcbprm)(2002)}. Thus, within the framework of the model in which the acceleration of electrons is caused by the electric field of Alfven oscillations in the ionospheric resonator, it is problematic to implement periodic sequences of S-bursts with a large number of radiation pulses. It should be noted that there are other structures, such as, for example, bursts with frequency splitting \citep{Flagg(fkl)(1976)} or bursts with a quasi-harmonic structure \citep{Panchenko(prrbzlskmfs)(2018)}, the appearance of which is difficult or impossible to explain within the framework of the electron cyclotron maser model.

In the work of \citet{Zaitsev(zzs)(1986)} another approach to the solution of the problem of occurrence of bursts of S-radiation and formation of quasiperiodic sequences of individual S-bursts in decameter radio emission associated with the satellite Io is demonstrated. This approach is free from the above-mentioned shortcomings of the ECM model. According to the model proposed in the work of \citet{Zaitsev(zzs)(1986)}, the source of radiation is nonequilibrium electrons with a velocity distribution function of the ''loss cone'' type generating plasma waves at the frequency of the upper hybrid resonance. Plasma waves are converted into electromagnetic radiation as a result of scattering on flows of suprathermal ions. Involvement of suprathermal ions in the model provides an increase in the scattering frequency, which makes it possible to overcome the ''stop band'', the width of which in a strong magnetic field  $\omega_{\rm B} \gg \omega_{\rm L}$  is of the order of  $\displaystyle {\frac{\omega_{\rm L}^2}{2\omega_{\rm B}}}$ ($\omega_{\rm L}$  and $\omega_{\rm B}$ are the plasma and cyclotron frequencies of electrons, respectively) and to form electromagnetic radiation corresponding to the extraordinary mode. The term ''stop band'' denotes the frequency interval between the maximum frequency of the plasma wave and the minimum frequency of the fast extraordinary electromagnetic mode. Under condition of  transverse propagation  these frequencies are respectively equal to the frequency of the upper hybrid resonance $\omega_{\rm p,max}=\omega_{\rm UH}=\sqrt{\omega_{\rm L}^2+\omega_{\rm B}^2}$ and the ''cutoff'' frequency of the extraordinary mode $\omega_{\rm e,min}=\omega_{\rm cut}=\sqrt{2\omega_{\rm L}^2+\omega_{\rm B}^2}$ \citep[see for details][]{Akhiezer(aapss)(1975)}. Quasiperiodic sequences of S-bursts appear as a result of the implementation of the pulsating conversion mode in the source at its nonlinear stage. In this case, the duration of the radiation pulse sequences is determined only by the lifetime of the flows of nonequilibrium electrons and ions, and the sequences themselves can contain a large number of radiation pulses. The negative frequency drift of an individual burst is due to the propagation effect. It is the result of the group delay of electromagnetic waves during their propagation from the generation region to the planet's magnetosphere. Note that the plasma model of the S-bursts source does not impose special requirements on the energy of the radiating particles. It is sufficient to have electrons whose velocities exceed the velocities of the electrons of the equilibrium plasma to avoid absorption of excited plasma waves by the particles of the equilibrium plasma due to Landau damping.

In this paper we investigate the possibility of implementing the above-mentioned plasma model of S-radiation generation for the case of S-bursts associated with Ganymede and estimate the plasma parameters in the generation region required for this implementation. Sections~2 and 3 provide the necessary information on the generation of plasma waves as a result of cone instability (section 2) and on the pulsating regime of conversion of excited plasma waves into electromagnetic waves. Section 4 provides an estimate of the plasma parameters at which the implementation of the plasma generation mechanism in the ionosphere of Jupiter and the occurrence of quasiperiodic negatively drifting in frequency bursts of decameter radiation with periods and frequency drift rates corresponding to observational data are possible.

\section{Excitation of plasma waves near the upper hybrid frequency}
\noindent

\begin{figure}
	\includegraphics[width=\columnwidth]{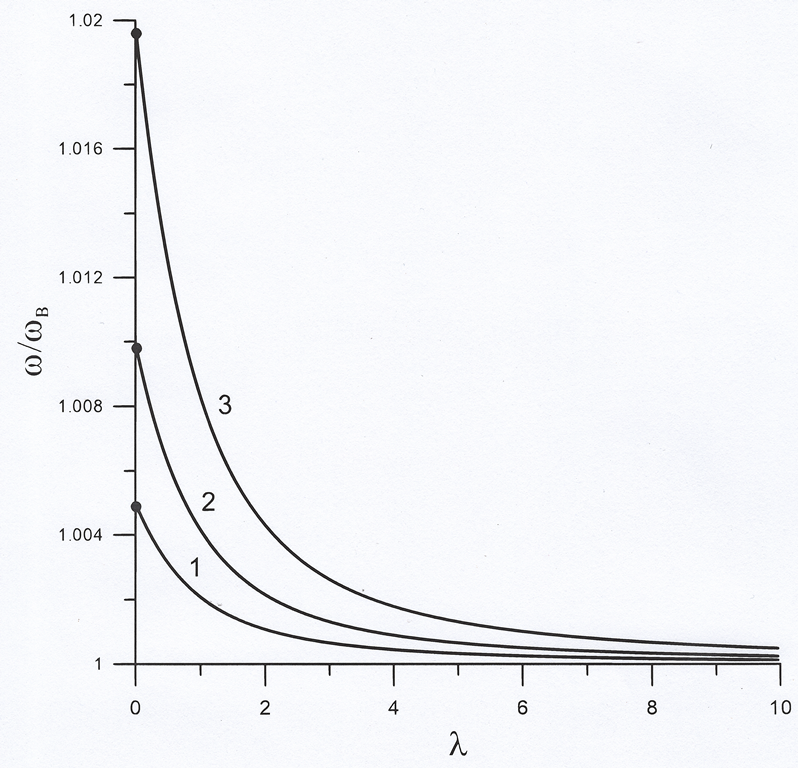}
    \caption{Dependence of the plasma wave frequency on the wave vector (the value  $\displaystyle \lambda=\frac{k_{\perp}^2 v_{\rm T0}^2}{\omega_{\rm B}^2} $) for different values $\displaystyle \frac{\omega_{\rm L}}{\omega_{\rm B}}$ of the ratio of the plasma frequency to the cyclotron frequency of electrons:  1)  $\displaystyle \frac{\omega_{\rm L}}{\omega_{\rm B}}=0.01 $; 2)  $\displaystyle \frac{\omega_{\rm L}}{\omega_{\rm B}}=0.02$; 3) $\displaystyle \frac{\omega_{\rm L}}{\omega_{\rm B}}=0.04$.  The filled circles on the ordinate axis indicate the frequency values equal to the frequency of the upper hybrid resonance $\omega_{\rm UH}$.}
    \label{fig2}
\end{figure}

In magnetoactive plasma, plasma waves propagating across the magnetic field are described by the dispersion relation \citep{Zaitsev(zzs)(1986)}
\begin{equation}
    \varepsilon_{\parallel}^{(0)}=1-\frac{2\omega_{\rm L}^2 I_{\rm 1}(\lambda)\exp (-\lambda)}{\lambda (\omega^2-\omega_{\rm B}^2)}=0
    \label{eps}
\end{equation}
where $I_{1}(\lambda)$ is the first-order Bessel function of the imaginary argument, $\displaystyle \lambda=\frac{k_\perp^2 v_{\rm T 0}^2}{\omega_{\rm B}^2 }$, $k_\perp$ is the component of the wave vector orthogonal to the magnetic field,  $\displaystyle v_{\rm T_0}=\sqrt{\frac{\kappa_{\rm B} T_0}{m}}$, $T_0$ is the temperature of equilibrium electrons, 
$\kappa_{\rm B}$ is the Boltzmann constant, $m$ is the electron mass. Relation (\ref{eps}) is also valid for waves propagating at angles other than orthogonal, provided that
\begin{equation}
|\omega-\omega_{\rm B}|\gg k_{\parallel}v_{\rm T_0},
\label{cond}
\end{equation}
at which the Landau absorption of waves is small \citep{Zheleznyakov(1996)}. In (\ref{cond}) $k_{\parallel}$ is the component of the wave vector along the magnetic field. The solution to equation (\ref{eps}) has the form
\begin{equation}
    \frac{\omega^2}{\omega_{\rm B}^2}=1+\frac{\omega_{\rm L}^2}{\omega_{\rm B}^2}\frac{2 I_{\rm 1}(\lambda)\exp (-\lambda)}{\lambda }.
    \label{disp}
\end{equation}
Fig.~\ref{fig2} shows the dispersion curves of the plasma wave for different values of the ratio $\displaystyle \frac{\omega_{\rm L}}{\omega_{\rm B}}$.

Let us assume that in the source of S-bursts, in addition to the equilibrium plasma described by the permittivity $\varepsilon_{\parallel}^{(0)}$, there is a small admixture,
 $N_{\rm e}\ll N_0$,  of hot nonequilibrium electrons, $T_{\rm e} \gg T_0$. Here $N_{\rm e}$ and $N_0$ are the concentrations of hot and equilibrium electrons, respectively, $T_{\rm e}$ is the temperature of hot electrons. The condition $T_{\rm e} \gg T_0$ ensures a small value of absorption of excited plasma waves by particles of the equilibrium component of the plasma. It is natural to assume that the hot component has a velocity distribution characteristic of trapped electrons in a magnetic field, for example, of the ''loss cone'' type.
\begin{equation}
f(v_{\parallel},v_{\perp})=\frac{N_{\rm e}v_{\perp}^2}{2(\sqrt{2\pi})^3v_{\rm e}^3}\exp -\left(\frac{v_{\parallel}^2+v_{\perp}^2}{2v_{\rm e}^2}\right)
\label{func}
\end{equation}
where $v_{\rm e}=\displaystyle \sqrt{\frac{\kappa_{\rm B}T_{\rm e}}{m}}$.
These electrons excite plasma waves with an increment
\begin{equation}
\gamma=-\frac{\rm Im\, \varepsilon_{\parallel}^{(1)}}{\left[\frac{\partial}{\partial\omega }\varepsilon_{\parallel}^{(0)} \right]_{\varepsilon_{\parallel}^{(0)}=0}}.
\label{gamma}
\end{equation}
Under the condition $\omega_{\rm B}\leq \omega_{\rm UH} \leq 2\omega_{\rm B}$  the imaginary part of the permittivity $\rm Im\, \varepsilon_{\parallel}^{(1)}$ caused by nonequilibrium electrons  (\ref{func}) has the form \citep{Zheleznyakov(zz)(1975a)}
\begin{eqnarray}
\rm Im\,\epsilon_{\parallel}^{(1)}  \simeq
\sqrt{\frac{\pi}{2}}\frac{\tilde{\omega}_{\rm L}^2\omega_{\rm B}}{k^2 k_{\parallel}v_{\rm e}^3}
 \exp (-Z^2)\times \\ \nonumber
\left[\delta(\omega)
\varphi (\xi)+(\delta (\omega)+1)\xi\varphi\,'(\xi)\right]
\label{im_epsilon}
\end{eqnarray}
where $\tilde{\omega}_{\rm e}=\sqrt{ \displaystyle {4\pi {\rm e}^2N_{\rm e} \over m}}$,
  $\varphi (\xi)=e^{-\xi} I_{1}(\xi)$,
$\xi = \displaystyle {k_{\perp}^2v_{\rm e}^2 \over \omega_{\rm B}^2}$,
$\delta (\omega) = \displaystyle {\omega - \omega_{\rm B}\over \omega_{\rm B}}$,
 $Z = \displaystyle {\omega - \omega_{\rm B}\over {\sqrt{2}k_{\parallel}v_{\rm e}}}$,
\begin{equation}
\left[\frac{\partial}{\partial\omega }\varepsilon_{\parallel}^{(0)} \right]_{\varepsilon_{\parallel}^{(0)}=0}=\frac{2\omega}{\omega^2-\omega_{\rm B}^2}.
\label{d_eps}
\end{equation}
From (\ref{gamma})-(\ref{d_eps}) for the optimal direction of excitation of the plasma wave, determined from the condition
\begin{equation}
\frac{|\omega-\omega_{\rm B}|}{k_{\parallel}^{\rm opt}v_{\rm e}}=1,
\label{opt}
\end{equation}
we obtain the following expression for the increment
\begin{eqnarray}
\gamma_{\rm opt}=-\frac{\pi}{2\sqrt{2e}}\frac{(2+\delta (\omega))}{(1+\delta (\omega))}\frac{N_{\rm e}}{N_0}\frac{\omega_{\rm L}^2}{\omega_{\rm B}}\frac{1}{\xi}\times \\ \nonumber
\left[\delta(\omega)
\varphi (\xi)+(\delta (\omega)+1)\xi\varphi\,'(\xi)\right].
\label{gamma_opt}
\end{eqnarray}

\begin{figure}
	\includegraphics[width=\columnwidth]{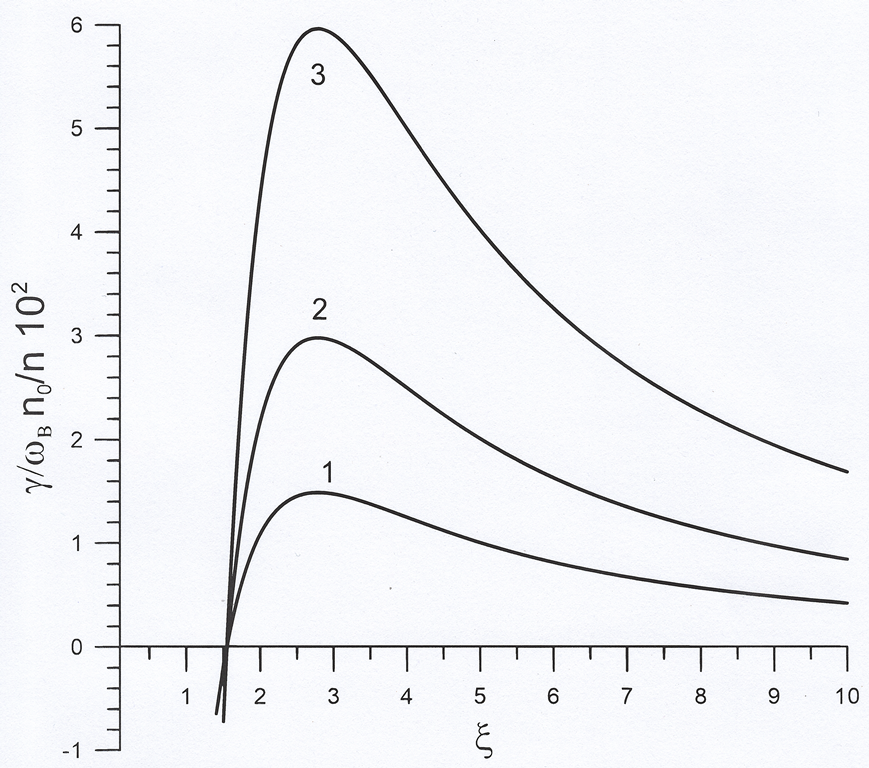}
    \caption{Dependence of the plasma wave increment on the wave vector (the value $\displaystyle \xi=\frac{k_{\perp}^2 v_{\rm e}^2}{\omega_{\rm B}^2} $ ) for different values of the ratio of the plasma frequency to the cyclotron frequency of electrons: 1)  $\displaystyle \frac{\omega_{\rm L}}{\omega_{\rm B}}=0.01 $; 2)  $\displaystyle \frac{\omega_{\rm L}}{\omega_{\rm B}}=0.02$; 3) $\displaystyle \frac{\omega_{\rm L}}{\omega_{\rm B}}=0.04$.}
    \label{fig3}
\end{figure}

Figure~\ref{fig3}  shows the dependence of the  increment of plasma waves on the wave vector for different values of the ratio of the plasma frequency to the cyclotron frequency of electrons.
It is evident from Figure~\ref{fig3} that the maximum increment is achieved at values of   $\xi\approx 3$. Taking into account $\displaystyle v_{\rm e}\gg v_{\rm T_0}$ and 
$\displaystyle {\lambda=\xi\frac{v_{\rm T_0}}{v_{\rm e}}}$, it follows from the dispersion curves presented in Figure~\ref{fig2} that the maximum increment is achieved for waves with frequencies close to the upper hybrid resonance frequency $\omega_{\rm max}\approx \omega_{\rm UH}$. Moreover, as estimates show, the value $\omega_{\rm max}$ changes little with a change in the energy of the radiating electrons in a wide energy range.

\section{Conversion of plasma waves into electromagnetic  extraordinary waves}
\noindent

Plasma waves cannot leave the source region, located in a sufficiently dense plasma in the upper ionosphere, into the more rarefied plasma of the magnetosphere and, even more so, into interplanetary space. Therefore, conversion of these waves into electromagnetic waves is necessary. Observations indicate that the decameter radio emission of Jupiter corresponds to an extraordinary mode \citep{Goldstein(gg)(1983)}. In order to convert plasma waves into an extraordinary mode, it is necessary to raise the radiation frequency by at least the value of the ''stop band'' from the frequency of the upper hybrid resonance to the minimum frequency of the extraordinary mode, $\Delta \omega \geq \omega_{\rm e,min}-\omega_{\rm UH}$. In a plasma with a strong magnetic field $\omega_{\rm B}\gg \omega_{\rm L}$, this condition can be written as follows 
$\displaystyle \Delta \omega \geq \frac{\omega_{\rm L}^2}{2\omega_{\rm B}}$. Following \citet{Zaitsev(zzs)(1985),Zaitsev(zs)(1988b)}, we assume that the conversion occurs as a result of scattering of plasma waves by a flow of suprathermal ions into the frequency region, where the refractive index of the extraordinary wave is small $n_{\rm e}\ll 1$. Flows of nonthermal ions are necessary because scattering on ions of the equilibrium plasma occurs with a decrease in the frequency of the scattered wave \citep{Zheleznyakov(1996)}. Due to this process, the energy of plasma waves can be transferred along the spectrum, as well as the conversion of plasma waves into electromagnetic waves corresponding to the ordinary mode. As can be seen below, these processes do not play a significant role in the formation of periodic sequences of bursts of electromagnetic radiation.

The processes of wave scattering on particles are described by a system of integro-differential equations of transfer, which in general are difficult to analyze \citep[see, in this connection,][]{Zheleznyakov(1996)}. As applied to the source of S-bursts of decameter radio emission from Jupiter, they can be simplified if we take into account the following circumstances. The high power of S-radiation and the small size of the source suggest a high energy density of radiation, both plasma and electromagnetic radiation, which indicates a large gain of these waves. In addition, the group velocity $v_{\rm gr}$ of excited plasma waves and extraordinary electromagnetic waves with a small refractive index $n_{\rm e} \ll 1$ is much less than the speed of light. All this  provides justification for neglecting spontaneous processes in comparison with induced ones, and also for neglecting the term $\displaystyle v_{\rm gr}\frac{\partial W}{\partial l}$, responsible for the change in wave energy due to its transfer in space. Due to the large group velocity of ordinary electromagnetic waves, which is of the order of the speed of light, the amplification of ordinary waves at the source sizes will be significantly less than the amplification of extraordinary and plasma waves, and it is possible to ignore the process of scattering into an ordinary wave. We will also neglect the process of scattering of plasma waves on ions of equilibrium plasma. Scattering of plasma waves by equilibrium ions leads to a ''pumping'' of waves along the spectrum toward lower frequency values. During the scattering process, this change in frequency is small $\Delta \omega \sim \displaystyle {\omega_{\rm B}\frac{v_{\rm Ti}}{v_{\rm e}}} $ ($v_{\rm Ti}$ is the thermal velocity of equilibrium ions). Estimates show that even taking into account the expansion of the frequency spectrum of plasma waves during scattering by ions of the equilibrium plasma, the width of this spectrum is significantly smaller than the change in frequency during conversion, which is of the order of $\sim \displaystyle {\omega_{\rm B}\frac{\omega_{\rm L}^2}{2\omega_{\rm B}^2}}$, and we can assume that the conversion is of an ''integral'' nature. In this approximation, it is possible to integrate the terms of the transport equation over the wave vector, resulting in a system of equations for the total energy density \citep[for details, see][]{Zheleznyakov(1996)}
\begin{eqnarray}
\frac{dW_{\rm p}}{dt} &= & 2(\gamma-\nu_{\rm p})W_{\rm p}-\eta W_{\rm p}W_{\rm e}, \nonumber \\
\frac{dW_{\rm e}}{dt} &= & -2\nu_{\rm e}W_{\rm e}+\eta W_{\rm p}W_{\rm e},
\label{syst}
\end{eqnarray}
where $W_{\rm p,e}$ is the  energy density of plasma (p) and extraordinary electromagnetic (e) waves in the source; $\eta$ is the coefficient of conversion of plasma wave into extraordinary electromagnetic wave, $\nu_{\rm p,e}$ is the absorption coefficient of plasma (p) and electromagnetic (e) waves, which coincides for plasma wave and extraordinary electromagnetic wave with small refractive index $n_{\rm e} \ll  1$ with effective frequency $\nu_{\rm eff}$ of electron-ion collisions in equilibrium plasma \citep{Ginzburg(gr)(1975)}
\begin{equation}
\nu_{\rm eff}=\sqrt{\frac{8\pi}{m}}\frac{e^2N_0}{(\kappa_{\rm B}T_0)^{3/2}}\ln\left(0.37\frac{\kappa_{\rm B}T_0}{e^2N_0^{1/3}}\right)\approx 50\frac{N_0}{T_0^{3/2}}.
\label{nu_eff}
\end{equation}

In the phase space $W_{\rm p},W_{\rm e}$, the solution of the system of equations (\ref{syst}) is a closed trajectory around a singular point of the ''center'' type 
\citep{Zaitsev(1971)}
\begin{equation}
W_{\rm p,0}=\frac{2\nu_{\rm eff}}{\eta}\,\,\,   W_{\rm e,0}=\frac{2(\gamma-\nu_{\rm eff})}{\eta}.
\label{center}
\end{equation}
\begin{figure}
	\includegraphics[width=\columnwidth]{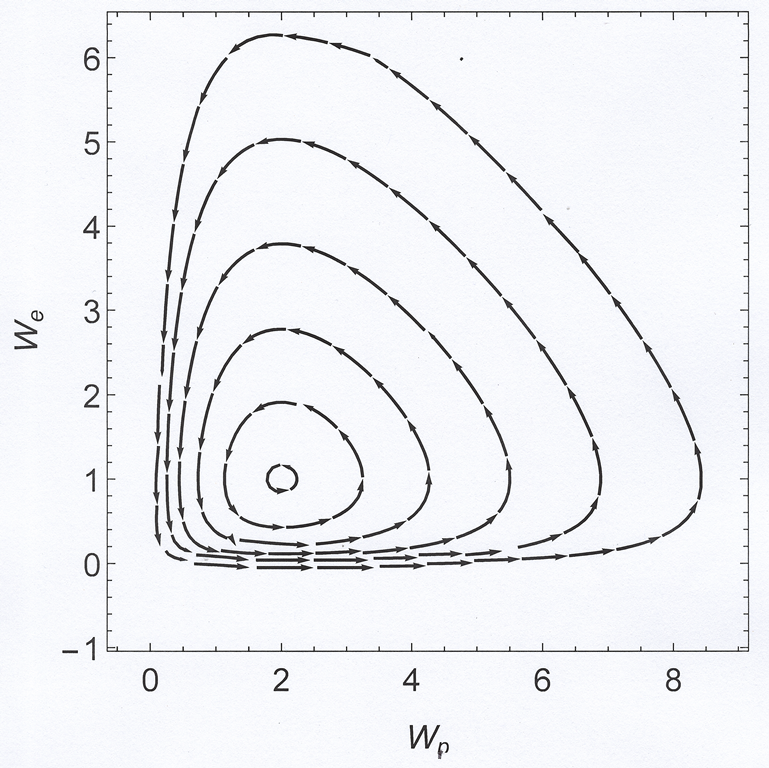}
    \caption{Trajectories of change in the process of conversion of the energy density of electromagnetic $W_{\rm e}$ and plasma $W_{\rm p}$ waves in the phase space $W_{\rm p},W_{\rm e}$ for the parameters of equation  (\ref{syst})  $\displaystyle {\frac{\nu}{\eta}=1}$; $\displaystyle {\frac{\gamma}{\eta}=1.5}$.}
    \label{fig4}
\end{figure}
The pulsation period for a large modulation value $|W_{\rm e}-W_{\rm e,0} |\sim W_{\rm e,0}$;  $|W_{\rm p}-W_{\rm p,0} |\sim W_{\rm p,0}$ and a large increment of plasma wave instability $\gamma \gg \nu_{\rm eff}$ is equal to
\begin{equation}
T_{\rm c}\approx \frac{1}{2\nu_{\rm eff}}\ln \left(\frac{W_{\rm e,0}}{W_{\rm e}(0)}\right)\approx  \frac{\Lambda}{2\nu_{\rm eff}},
\label{T_big}
\end{equation}
where $W_{\rm e}(0)$ is the energy density at the initial moment of time, $\Lambda$ is a value approximately equal to the Coulomb logarithm. In this case, the pulse duration turns out to be significantly less than the distance between the pulses.
For small deviations from the equilibrium state $|W_{\rm e}-W_{\rm e,0} |\ll W_{\rm e,0}$;  $|W_{\rm p}-W_{\rm p,0} |\ll W_{\rm p,0}$ and a large increment of plasma wave instability $\gamma \gg \nu_{\rm eff}$  the period is \citep{Zaitsev(1970)}
\begin{equation}
T_{\rm c}\approx \frac{\pi}{\sqrt{\gamma \nu_{\rm eff}}}.
\label{T_small}
\end{equation}
In this case, the duration of individual pulses of electromagnetic radiation coincides in order of magnitude with the period $T_{\rm c}$.

Note that the averaged value of the wave energy density over the pulsation period does not depend on the modulation depth and is equal to the corresponding equilibrium value

Thus, as a result of the conversion, a periodic sequence of electromagnetic wave pulses corresponding to the extraordinary mode with a small refractive index $n_{\rm e} \ll 1$ arises.

The negative frequency drift is caused by the group delay effect when electromagnetic radiation leaves the generation region and enters the region of Jupiter's magnetosphere, where the refractive index of electromagnetic waves is close to unity. The time it takes for a signal to travel the path $l$, on which the refractive index changes from $n_{\rm e} \ll 1$ to $n_{\rm e}\approx 1$, is determined by the integral $t(\omega)\approx \int \displaystyle {\frac{dl}{v_{\rm gr}(\omega)}}$,  where $v_{\rm gr}(\omega)$, where  is the group velocity and $l$ is the coordinate along the wave propagation trajectory. The magnitude of the frequency drift of the burst observed on the frequency-time diagram and caused by the dependence of time on frequency $t(\omega)$ can be estimated by the formula
\begin{equation}
\frac{1}{\omega}\frac{d\omega}{dt}=-\frac{c}{L_{\rm N}}\left(\frac{\omega_{\rm L}}{\omega_{\rm B}}\right)^4\zeta
\label{drift}
\end{equation}
where $L_{\rm N}$ is the characteristic distance at which the refractive index changes from values $n_{\rm e}\ll 1$ to $n_{\rm e} \approx 1$, $\zeta \sim (0.1-1)$ is a parameter depending on the nature of the change in the refractive index $n_{\rm e}$ along the trajectory of radiation propagation, $c$ is the speed of light.

\section{Formation of S-burst sequences of frequency-drifting decameter radio emission associated with Ganymede}
\noindent

Let us consider the possibility of using the plasma model to interpret the observed periodic sequences of frequency-drifting S-bursts of decameter radio emission related to Ganymede and presented in Fig.~\ref{fig1}. To do this, we will estimate the plasma parameters in the radiation source necessary for the formation of such sequences, and compare the obtained results with known data on the magnetosphere and ionosphere of Jupiter.

As follows from the plasma model and relation (\ref{drift}), the frequency drift is caused by the group delay of electromagnetic waves, and its speed depends significantly on the ratio of the plasma frequency to the electron gyrofrequency. Assuming that the characteristic scale $L_{\rm n}$ in (\ref{drift}) is of the order of the size of the ionospheric source $L_{\rm n}\simeq 4\times 10^7 \mbox{ cm}$ \citep[see, in this connection,][]{Grodent(gbrgjnc)(2009)}, we obtain the following estimates of the ratio of the plasma frequency to the electron gyrofrequency. For the burst shown in Fig.~\ref{fig1}a, where the frequency drift rate is $-16.9  \mbox{ MHz/s}$, we obtain from (\ref{drift}) the ratio of the plasma frequency to the gyrofrequency of electrons in the generation region $ \displaystyle {\frac{\omega_{\rm L}}{\omega_{\rm B}}\simeq 2\times10^{-1}}$, and for the burst shown in Fig.~\ref{fig1}b, where the drift rate is $-3.4 \mbox{ MHz/s}$, the ratio $\displaystyle {\frac{\omega_{\rm L}}{\omega_{\rm B}}\simeq 10^{-1}}$. According to the model, the electromagnetic waves generated as a result of plasma wave scattering have frequencies close to the ''cutoff'' frequency for the extraordinary mode $\omega_{\rm cut}$, which, under the conditions of the discussed S-burst source, differs little from the electron gyrofrequency $\omega_{\rm e}\approx \omega_{\rm cut}=\sqrt{\omega_{\rm B}^2+2\omega_{\rm L}^2}\approx \displaystyle {\omega_{\rm B}+\frac{\omega_{\rm L}^2}{\omega_{\rm B}}}$. Assuming $\omega_{\rm e}=2\pi\times 15 \mbox{ MHz}$ \citep{Mauduit(mzlh)(2023)},
we obtain the following estimate of the equilibrium plasma concentration in the generation region $N_0 \simeq (0,5-1)\times 10^5 \mbox{ cm$^{-3}$}$, which is in agreement with the known data on the plasma concentration in the ionosphere of Jupiter \citep{Hinson(hfksth)(1997)}.

The required concentration of electrons exciting plasma waves can be estimated from the expression for the repetition period of S-bursts (\ref{T_small}). According to \citet{Mauduit(mzlh)(2023)}, the pulse repetition period in the quasi-periodic radiation sequences shown in Fig.!\ref{fig1}  is 31~ms and 38~ms, respectively. Moreover, visually, the distance between pulses coincides in order of magnitude with the pulse repetition period. Assuming the plasma temperature in the Jupiter ionosphere to be  $T_0\approx 1500 \mbox{ K}$ \citep{Hinson(hfksth)(1997)} and the equilibrium plasma concentration in the generation region to be $10^5  \mbox{ cm$^{-3}$}$ and $5\times 10^4 \mbox{ cm$^{-3}$}$, respectively, we obtain the following estimate of the concentration of energetic electrons in the flow: $ N_{\rm e} \approx 10^2 \mbox{ cm$^{-3}$}$.

As shown in Section~2, plasma waves are mainly excited at frequencies close to the upper hybrid resonance frequency $\omega_{\rm p}\approx \omega_{\rm UH}=\sqrt{\omega_{\rm B}^2+\omega_{\rm L}^2} \approx \omega_{\rm B}+ \displaystyle {\frac{ \omega_{\rm L}^2}{2\omega_{\rm B}}}$. When these waves are scattered by an ion with a velocity of $v_{\rm i}$, the frequency of the resulting electromagnetic wave $\omega_{\rm e}$ is determined from the conservation law
\begin{equation}
\omega_{\rm e}=\omega_{\rm p}+(\vec{k}_{\rm e}-\vec{k}_{\rm p})\vec{v}_{\rm i}\approx \omega_{\rm p}+k_{\rm p}v_{\rm i\perp},
\label{synchr}
\end{equation}
where $k_{\rm p}\approx \displaystyle {\frac{\omega_{\rm p}}{v_{\rm e}}\approx \frac{\omega_{\rm B}}{v_{\rm e}}}$ and $k_{\rm e}=\displaystyle {\frac{\omega_{\rm e}}{c}n_{\rm e}}$ are the wave vectors of the excited plasma wave and the converted electromagnetic wave, respectively. In (\ref{synchr}) it is taken into account that $k_{\rm p}\gg k_{\rm e}$ and the plasma waves are excited at small angles to the direction orthogonal to the magnetic field in the source. Thus, the conversion of plasma waves into electromagnetic waves, which is accompanied by an increase in frequency by $\Delta \omega =\omega_{\rm e}-\omega_{\rm p}\approx \displaystyle {\frac{\omega_{\rm L}^2}{2\omega_{\rm B}}}$, is possible provided that the velocity of the scattering ions exceeds $v_{\rm i\perp}\geq \displaystyle {\frac{\Delta \omega}{\omega_{\rm B}}v_{\rm e}} \approx \displaystyle { \frac{\omega_{\rm L}^2}{2\omega_{\rm B}^2}v_{\rm e}}$. Provided that scattering occurs on protons and assuming that $\displaystyle {\frac{\omega_{\rm L}^2}{\omega_{\rm B}^2}} \approx 4\times 10^{-2}$, which corresponds to the burst shown in Fig.~\ref{fig1}a, the specified condition in energy units takes the form  $\varepsilon_{\rm i} \geq 40\varepsilon_{\rm e}$,  where $\varepsilon_{\rm e,i}$ is the characteristic energy of the electrons (e) emitting plasma waves and the ions (i) scattering these waves. For a burst with smaller drift velocities in absolute value (Fig.~\ref{fig1}b), the energy of the scattering ions can be smaller $\varepsilon_{\rm i}\geq 10\varepsilon_{\rm e}$.

In the framework of the model discussed in the paper, the energy of radiating electrons is a free parameter. The  electrons from a wide range of energies can effectively excite plasma waves. Measurements with Juno showed that the electrons  propagating in the Alfven tube of Ganymede  have a wide energy spectrum, extending from several tens of electronvolts to several tens of kiloelectronvolts with an increase in the flux towards lower energies. Moreover, electrons accelerated directly in the tube were observed at energies 
$\varepsilon_{\rm e}\geq 70 \mbox{ eV}$ \citep{Rabia(r19)(2024)}. Assuming, for definiteness, that this energy is the energy of radiating electrons 
$\varepsilon_{\rm e}=70 \mbox{ eV}$, we obtain the following estimate of the energy of ions required for conversion $\varepsilon_{\rm i} \simeq (0.7-2.8) \mbox{ keV}$.

\section{Conclusion}

 On the example of radiation bursts  presented in the paper by \citet{Mauduit(mzlh)(2023)}, the possibility of implementing a plasma mechanism for both the generation of the S-bursts  related to Ganymede and  the formation of quasi-periodic sequences of frequency-drifting emission pulses is considered. This mechanism, which was first proposed by \citet{Zaitsev(zzs)(1985)} to explain narrow-band quasi-periodic pulse sequences of Io-associated S-radiation, is a two-stage process. At the first stage, plasma waves are excited by energetic electrons with a nonequilibrium distribution function in transverse velocities. Plasma waves are excited at frequencies near the upper hybrid resonance. Within the framework of the discussed model, the energy of electrons exciting plasma waves is a free parameter and has a weak effect on the excitation efficiency. In the proposed model, an energy of $\sim 70 \mbox{ eV}$ was taken for estimates, corresponding to the energy of electrons accelerated directly in the Alfven tube of Ganymede \citep{Rabia(r19)(2024)}.
 
 Plasma waves cannot go beyond the planet's magnetosphere. Conversion of these waves into electromagnetic radiation is necessary. This occurs at the second stage of the generation process. According to the model, as a result of scattering of plasma waves on energetic ion flows, these waves are converted into extraordinary electromagnetic waves with frequencies close to the ''cutoff'' frequency of the extraordinary mode $\omega_{\rm cut}\approx  \omega_{\rm B}+\displaystyle {\frac{\omega_{\rm L}^2}{\omega_{\rm B}}}$, where the refractive index of the extraordinary wave is small $n_{\rm e} \ll 1$. In the process of scattering, due to the energy of ions, the frequency of scattered waves increases, which is necessary to overcome the so-called ''stop band'', the width of which is equal to $\Delta \omega \approx \displaystyle {\frac{\omega_{\rm L}^2}{2\omega_{\rm B}}}$. The ion energy required for this depends, in addition to the ratio $\displaystyle {\frac{\omega_{\rm L}}{\omega_{\rm B}}}$, on the energy of the electrons exciting the plasma waves $\varepsilon_{\rm e}$. For the radiation bursts discussed in the paper, the necessary conversion can be provided by ions with energy $\varepsilon_{\rm i} \simeq (0.7-2.8) \mbox{ keV}$.
 
The effect of group delay of electromagnetic waves with a small refractive index shows itself in the dynamic spectrum of radiation as a negative frequency drift. Its value depends significantly on the concentration of the main plasma in the source via the ratio $\displaystyle {\frac{\omega_{\rm L}}{\omega_{\rm B}}}$. Estimates show that the observed frequency drift velocities are realized at a plasma concentration of $N \sim 10^5 \mbox{ cm$^{-3}$}$. Moreover, lower absolute drift velocities are realized in a source with a lower plasma concentration. Thus, for the radiation bursts shown in Fig.~\ref{fig1}, the concentration  $N_0 \simeq 10^5  \mbox{ cm$^{-3}$}$  ($\displaystyle {\frac{\omega_{\rm L}}{\omega_{\rm B}}\simeq 2\times 10^{-1}}$) corresponds to a drift velocity of $\simeq -17 \mbox{ MHz/s}$, and the concentration $N_0 \simeq 5\times 10^4  \mbox{ cm$^{-3}$}$  ($\displaystyle {\frac{\omega_{\rm L}}{\omega_{\rm B}}\simeq  10^{-1}}$) corresponds to a drift velocity of $-3.4 \mbox{ MHz/s} $.

The implementation of the pulsating conversion mode in the source provides the appearance of quasi-periodic pulse sequences. The radiation pulsation period depends on the parameters of the main plasma in the source, the temperature $T_0$ and the concentration $N_0$, as well as on the concentration of energetic nonequilibrium electrons $N_{\rm e}$. We found that at $T_0=1500$~K and $N_0=(0.5-1)\times 10^5 \mbox{ cm$^{-3}$}$, energetic electrons with a concentration of $N_{\rm e}\simeq 10^2  \mbox{ cm$^{-3}$}$ provide the observed radiation pulse repetition periods of $31 \mbox{ ms}$  and $38 \mbox{ ms}$.

\citet{Mauduit(mzlh)(2023)} note that bursts with lower absolute drift velocities are generally less intense than fast-drifting bursts. In the plasma model, this effect is a consequence of the conversion of plasma waves into electromagnetic waves corresponding to the extraordinary mode with a low refractive index. According to the model, the frequency drift rate depends on the value of the  ratio $\displaystyle {\frac{\omega_{\rm L}}{\omega_{\rm B}}}$  at the source: the smaller the ratio, the lower the drift velocity. On the other hand, the ratio $\displaystyle {\frac{\omega_{\rm L}}{\omega_{\rm B}}}$ determines the width of the ''stop band'' $\Delta \omega \simeq \displaystyle {\frac{\omega_{\rm L}}{\omega_{\rm B}}}$, which must be overcome in the conversion process. The smaller the ratio $\displaystyle {\frac{\omega_{\rm L}}{\omega_{\rm B}}}$, the narrower the ''stop band'' and the lower the ''cutoff'' frequency of the extraordinary mode $\omega_{\rm cut}\simeq \omega_{\rm B} +\displaystyle {\frac{\omega_{\rm L}^2}{\omega_{\rm B}}}$.
This means that for the same energy of scattering ions, at lower values of  $\displaystyle {\frac{\omega_{\rm L}}{\omega_{\rm B}}}$, the conversion will occur in the frequency range where the refractive index is higher. According to the dispersion curve for the extraordinary mode, the farther the wave frequency is from the ''cutoff'' frequency, the higher the refractive index. The spectral radiation flux $F_{\omega}$ observed at some distance from the source is inversely proportional to the third power of the refractive index $F_\omega \propto \displaystyle {\frac{W_{\rm e}}{n_{\rm e}^3}}$, which is due to the dependence of the spectral radiation flux on the phase volume in the space of wave numbers \citep{Zaitsev(zzs)(1986)}. 1986). It follows that for the same value of the energy density of electromagnetic waves in the source $W_{\rm e}$, with a decrease in the ratio $\displaystyle {\frac{\omega_{\rm L}}{\omega_{\rm B}}}$  and, as a consequence, an increase in the refractive index, the spectral radiation flux $F_\omega$ recorded from the source will be lower.

In general, our study shows that the plasma generation mechanism is capable of generating quasi-periodic sequences of negatively drifting pulses of decameter radiation associated with Ganymede, with plasma parameters in the source being in good agreement with known data on the ionosphere of Jupiter. The rate of the recorded frequency drift can be used for radio astronomical diagnostics of the plasma concentration in the upper ionosphere of the planet in the region of decameter radio emission generation, which is subject to the influence of energetic electron and ion flows. The question of the formation of ionospheric plasma parameters in the active region of the ionosphere requires special consideration and is beyond the scope of this work.

\section*{Acknowledgements}
The study was carried out with financial support from the Russian Science Foundation (grant 25-22-00238).


\label{lastpage}
\end{document}